# Bilateral collaboration between Mexico and the United Kingdom for the construction of technical equipment (CHARM) for the Large Millimeter Telescope (LMT/GTM)

*Paulina Carmona Rodríguez[1], María de la Paz Ramos-Lara[2]*

**Abstract.** The Large Millimeter Telescope "Alfonso Serrano" (LMT), is the world's largest millimeter radio telescope and was founded in 2006. This radio telescope is the final product of a collaboration agreement between Mexico and the United States in the 1990s. It is located on top of an extinct volcano in Mexico at an altitude of 4600 meters above sea level. In 2018, the University of Manchester and the Rutherford Appleton Laboratory signed an agreement with the Instituto Nacional de Astrofísica Óptica y Electrónica (INAOE) to train Mexican astronomers in high-frequency radio receiver construction techniques by designing an innovative device called Collaborative Heterodyne Amplifier Receiver for Mexico (CHARM) designed to work at a frequency of 345 GHz. The research team, composed of British and Mexican technicians and scientists, installed CHARM at the LMT in 2019 and began testing the equipment until the COVID-19 pandemic shut it down in March 2020. This paper describes the collaboration process between Mexico and the United Kingdom, facilitated by a British institution dedicated to supporting scientific projects in developing countries, the Global Challenges Research Fund (GCRF).

[1] Facultad de Ciencias, Universidad Nacional Autónoma de México, carmonapaulina@ciencias.unam.mx
[2] Centro de Investigaciones Interdisciplinarias en Ciencias y Humanidades, Universidad Nacional Autónoma de México, ramoslm@unam.mx

**Background of the Large Millimeter Telescope "Alfonso Serrano" in Puebla, Mexico**

The geographical and meteorological conditions of certain developing countries have proven to be excellent for astronomical observations in the visible, radio, and infrared spectrums. As a result, first world countries have sought collaboration agreements with some of these nations to establish large telescopes in their territories. This is exemplified by the Atacama Large Millimeter Array (ALMA) in Chile and the Arecibo telescope in Puerto Rico, among others.

The initiative to build the Large Millimeter Telescope in Mexico began in 1988, when Dr. Paul Goldsmith, the director of the University of Massachusetts at Amherst (UMass-Amherst) in the United States, proposed the establishment of a bilateral collaboration agreement with Dr. Alfonso Serrano Pérez-Grovas (1950-2011), who at the time was the director of the Institute of Astronomy of the Universidad Nacional Autónoma de Mexico (UNAM),[1] to establish a bilateral collaboration agreement. The goal was to build a radio telescope with an antenna that would be the largest of its kind in the world, with a diameter of 50 meters. At that time, they were competing against the 30-meter diameter German-French-Spanish radio telescopes (IRAM in Pico Veleta, Spain), as well as the Nobeyama 45-meter diameter Japanese radio telescopes.[2]

The progress was significantly slow because this astronomical project was considered too costly for Mexico, given the country's economic situation, and was, in fact, the most expensive project in the history of science in Mexico. The initial cost estimate exceeded the entire budget of the UNAM, so it required funding from the federal government, U.S. institutions, and international organizations. In 1994, the project was approved with the involvement of UMass, the Instituto Nacional de Astrofísica Óptica y

Electrónica (INAOE),[3] which was established in 1971 in Puebla (Mexico), the Consejo Nacional de Ciencia y Tecnología (CONACYT, Mexico),[4] the Ministry of Finance and Public Credit of Mexico, and the World Bank through the Program for the Support of Science in Mexico (PACIME). The construction was expected to take between 10 and 12 years.[5]

The Large Millimeter Telescope (Gran Telescopio Milimétrico, GTM) Alfonso Serrano was designed through institutional scientific cooperation between Mexico (70%) and the United States (30%). The program aimed to enhance the scientific and technological capabilities of both countries by exchanging people, ideas, skills, experiences, and information.[6] The Large Millimeter Telescope (LMT) has an antenna that is 50 meters in diameter and can detect millimeter waves ranging from 0.85 to 4 millimeters. This range is substantial because a significant portion of the matter in the Universe is very cold (between 10 and 60 kelvins), and at least half of the radiated energy consists of waves ranging from millimeters to infrared. Millimeter light remains a mystery that can be unveiled with the help of large telescopes like the LMT. The LMT is one of the largest radio telescopes in the world, and it was proposed to explore the formation and evolution of galaxy clusters, galaxies, nearby stars, the central black hole of our galaxy, the formation of stars in the Milky Way, and the planets in our solar system.[7]

Due to the extraordinary sensitivity of the LTM, scientists expect to study the formation of complex organic molecules in molecular clouds, envelopes, circumstellar clouds, and comets. Also, the cosmochemistry that led to the creation of life and the analysis of the solar nebula in which the Sun and Earth formed.[8] The LMT's current objectives are to study the origin and evolution of structures on the largest cosmic scales. The following phenomena are expected to be investigated: the cosmic background radiation, thousands of galaxies forming in the early universe, stars in the process of formation, active galactic nuclei and supermassive black holes, objects exhibiting gamma-ray bursts, the distribution of interstellar gas and cold dust around forming stars in nearby and distant galaxies, gaseous disks with proto-planetary systems, small objects in solar systems, small molecules in planetary atmospheres, interplanetary space and the interstellar medium, comets to analyze their chemical and physical composition and understand how our solar system was formed.[9]

After conducting extensive analysis of various regions in Mexico, the technical and scientific staff of INAOE determined that the optimal location for the Large Millimeter Telescope installation would be the summit of Tliltépetl, an extinct volcano situated within the Pico de Orizaba National Park in the state of Puebla. This site, positioned at an altitude of 4600 meters (height above sea level) and approximately 100 kilometers from the city of Puebla, features favorable characteristics, including low water vapor content in the sky, minimal temperature fluctuations (5 degrees between seasons), moderate winds, and a manageable slope that facilitated the construction of a road for summit access. Furthermore, the site's proximity to the Mexico City airport, approximately 4 hours by highway, enhances logistical accessibility.[10]

In November 2006, the President of Mexico, Vicente Fox Quesada, inaugurated the Large Millimeter Telescope (LMT). This project is the most significant and expensive scientific endeavor in Mexico's history, with a 120 million-dollar budget. This is part of the CONACYT National Laboratories network. This radio telescope utilizes heterodyne receivers to detect and measure radiation with the highest possible accuracy. UMass has developed an instrument called SEQUOIA, which functions in the 80 to 120 GHz range. However, the antenna was expected to operate on a broader frequency range of 50 to 350

GHz. The telescope also includes precision detectors such as the Bolocam, which consists of a 144-bolometer array. Bolometers serve as radiation detectors, particularly adept at capturing emissions from outer space. It would be helpful to elaborate on the significance of studying molecular clouds and the formation of new stars, highlighting the broader implications for our understanding of the universe, exemplifying just one of many observable phenomena. Bolometers are devices designed to measure the intensity of incident electromagnetic radiation, especially in the infrared region. They are detectors for radiation coming from cold objects in the Universe presents in molecular clouds where new stars are being born. The molecular clouds are just one example of the many observable phenomena that will aid in expanding our comprehension of the Universe.[11]

Additional instruments, such as the Aztronomical Thermal Emission Camera (AZTEC), have been developed and constructed for the Large Millimeter Telescope. This specialized continuum camera has been used as a visiting instrument in other telescopes, including the 15-meter Anglo-Dutch-Canadian James Clerk Maxwell telescope. During 2005 and 2006, it was used to create a catalog of galactic sources. Additionally, it has been utilized in the Atacama Telescope in Chile.[12]

The Large Millimeter Telescope can operate at a wide range of frequencies. As a result, researchers from the University of Manchester and the Rutherford Appleton Laboratory developed a heterodyne detector known as the Collaborative Heterodyne Amplifier Receiver for Mexico (CHARM), which operates at 345 GHz. They were able to take advantage of support programs for developing countries funded by the Global Challenges Research Fund (GCRF), which we will discuss in the final section. Additionally, we will now provide some background information on these British institutions.

**Background of the University of Manchester, the Rutherford Appleton Laboratory, and the Global Research Fund**

The beginning of radio astronomy in England was stimulated by the discovery of radio waves emitted by the Sun. This discovery took place during World War II when anti-aircraft radar stations experienced significant signal interference. Initially, it was suspected to be caused by enemy activity. However, British physicist James Stanley Hey (1909-2000) determined that the signals were originating from the Sun and were connected to sunspot activity. These sunspots were confirmed by the Royal Observatory, Greenwich. After the war, systematic studies of solar radiation began. At Cambridge, a radar research scientist, the British astronomer Martin Ryle (1918-1984), used military equipment to build sensitive receivers that could measure solar radiation continuously. He also began to detect this radiation in stars and galaxies, which could be used to support cosmological theories.[13]

New technologies emerged due to the need for more powerful and sensitive equipment, such as radio transmitters, sensitive receivers, and radio antennas. At that time, the main leaders in radio astronomy were Martin Ryle at Cambridge, the British physicist and radio astronomer Bernard Lovell (1913-2012) at Manchester, and the Australian scientist and radio astronomer Joseph Pawsey (1908-1962) at Sydney. Lovell established the Jodrell Bank Observatory in Cheshire in 1945, the first to have radio telescopes. He did so through the Jodrell Bank Astrophysical Centre of the University of Manchester to study cosmic rays. On the other hand, Ryle focused on planning the construction of the Mullard Radio Astronomical Observatory, which was inaugurated in 1957. The research conducted with this type of radio telescope led to the development of the millimeter and submillimeter radio astronomy field. Ryle and Antony Hewish were awarded the 1974 Nobel Prize in

Physics for their work in radio astronomy and pulsars, making them the first astronomers to receive this award.[14]

In the United Kingdom, there are several institutions dedicated to funding scientific research projects. The United Kingdom Research and Innovation (UKRI) aims to invest in science and technology to improve people's lives, contribute to economic growth, generate jobs, and provide high-quality public services. The organization encompasses various research areas, from digital technologies to clean energy. The Science and Technology Facilities Council (STFC) is a subdivision of UKRI, and the Rutherford Appleton Laboratory (RAL) was established on the Chilton campus in Oxfordshire. RAL includes a space division called RAL Space, which oversees the research project The Accelerator Science and Technology Centre (ASTEC), initiated in 2001. ASTEC has a dedicated research team known as the Millimeter Wave Technology Group (MWTG), which focuses on developing advanced technology in radio astronomy and atmospheric sciences.[15]

The United Kingdom manages a program called the Global Challenges Research Fund (GCRF), which was created by the United Nations (UN). It supports various developing countries to promote scientific and technological research projects that can help improve their quality of life. As part of this program, there is the Newton Fund, of which Mexico is a partner. This partnership established a fund of approximately 12 million pounds for three years so that both countries can develop bilateral and multilateral programs focused on research and innovation.[16] In 2018, the University of Manchester and the Rutherford Appleton Laboratory signed an agreement with the Instituto Nacional de Astrofísica Óptica y Electrónica (INAOE) to design and build the CHARM device to be tested at the Mexican Large Millimeter Telescope.

**Bilateral agreement between Mexico and the United Kingdom for the design and construction of the CHARM detector for the Large Millimeter Telescope.**

In 2018, research collaboration was established among the University of Manchester, the Rutherford Appleton Laboratory (RAL), and the National Institute of Astrophysics, Optics, and Electronics (INAOE) to develop and install the CHARM detector in the Large Millimeter Telescope (LMT). This section gives insight into the collaboration process through interviews conducted by the Mexican physicist Paulina Carmona Rodríguez[17] with academic professionals from British and Mexican institutions involved in building CHARM.[18]

The International Symposium on Space Terahertz Technology (ISSTT) was established in 1990 to facilitate the exchange of information on emerging applications in astrophysics, planetary science, earth science, and remote sensing.[19] During the 19th ISSTT, a paper titled "Development of a 340-GHz Sub-Harmonic Image Rejection Mixer Using Planar Schottky Diodes" was presented by Bertrand Thomas, Simon Rea, Brian Moyna, and Dave Mathson. The collaborative effort involved researchers from RAL, NASA-JPL, and EADS-ASTRIUM Ltd, resulting in the development of the SHIRM (Sub-Harmonic Image Rejection Mixer) device. According to the article, observations of the Earth's atmosphere in the limb using spaceborne submillimeter waves can provide information on the global distribution of key molecular species in the upper troposphere and lower stratosphere. In radio astronomy, this information helps separate sidebands (from receivers). In 2014, the SHIRM space instrument, developed by RAL and other companies with support from the UK Centre for Earth Observation Instrumentation, was rigorously tested at the Sphinx Observatory on the Jungfraujoch, Switzerland, showcasing its potential in space instrument design.[20]

After years of not using SHIRM, Dr. Gary Fuller, a British astrophysicist at the University of Manchester, came up with the idea of using SHIRM technology to develop a new detector for the Large Millimeter Telescope (LMT). Fuller reached out to Dr. Brian Ellison, head of the millimeter wave group at the RAL, and suggested using the Global Challenges Research Fund program to create a new device tailored to the characteristics of the LMT and based on SHIRM technology. During the process, Mexican astronomers would be trained in heterodyne terahertz (THz) radio technology required to build this type of receivers; these techniques have both scientific and industrial application.[21] Fuller also presented the proposal to Dr. David Hughes, Director of the Large Millimeter Telescope, who agreed to collaborate on this joint project between Mexico and the United Kingdom. The project received £74,214 in funding from the GCRF between January 2018 and March 2022.

The team was assembled, and the Mexican scientist Dr. Edgar Colín was appointed as the representative of the LMT due to his expertise in electronics and astronomical instrumentation. The United Kingdom team was composed of Drs. Nart Daghestani, Eimear Gallagher and someone else. Gallagher joined the RAL in 2018 for her placement year before graduating from Nottingham Trent University. It is important to note that the United Kingdom has a "placement year" program, which involves working for a year in a company related to one's studies before graduation. The three members of the British group traveled to Mexico in October 2018 to visit the facilities of the LMT, where they met with Colín and the technical and scientific staff of INAOE. This visit was important for learning about the radio telescope's characteristics and for measuring the spaces where the new device called CHARM would be installed.

After their initial visit to the LMT, the English team and Colín returned to the UK. Colín spent several months at the RAL for technical training. Following working sessions in the UK led by Fuller, it was decided that CHARM would operate at a frequency of 345 GHz. CHARM was ready a few months later, and the legal procedures to import the technical equipment to Mexico began. Due to security issues on the road leading to the LMT, the technicians and equipment were escorted by state police.[22] The equipment was ready in the summer of 2019, just in time for the optimal observation months in the LMT, between November and January. The testing months proved fruitful and demonstrated the feasibility of using CHARM. The first results were presented at the 45th International Conference on Infrared, Millimeter and Terahertz the following year. The work entitled "Deployment of a fully integrated 360 GHz Schottky Diode Based" was presented by N. S. Daghestani, F. Cahill, A. Obeed, E. Gallagher, E. Colin-Beltrán, D. Sánchez-Arguelles, S. Kurtz, D. Hughes, G. A. Fuller, M. Dunstan, S. Parkes and B. N. Ellison. They discussed technical details of the detector's operation at 360 THz for observing first light with the LMT in the sub-millimeter-wave range, and their expectations to extend the range of observations.[23]

Authors Colín-Beltrán, Fuller, Ellison, Kurtz, Hughes, Daghestani, Cahill, Gallagher, Obeed, Sánchez-Argüelles, and Parkes published the results in the Proceedings of SPIE in 2020 titled "Charm: a room-temperature 345 GHz receiver for the Large Millimeter Telescope." In the paper, they reported on tests performed in mid-February 2020 with CHARM on the LMT on two occasions. During the first test, the antenna was directed towards the lunar disk, but the weather conditions were not favorable. On the second observing night, the weather conditions were better, and the researchers were able to make observations of the planets Jupiter and Venus. However, their sky tracking time was not sufficient to properly calibrate the devices. The researchers concluded that future observations could be directed towards star-forming regions, one of its primary objectives.[24]

Despite the health crisis caused by the COVID-19 pandemic and the temporary suspension of activities at the LMT, Colín had the opportunity to make observations using the CHARM in the LMT and successfully got results from these observations, which proved the correct functionality of the instrument. However, CHARM could not return to the UK within the official time frame due to the pandemic, and it is still seeking its next adventure, either in Mexico or the UK.

The technology behind CHARM would greatly benefit the automotive industry in Mexico, one of the country's biggest industries. It also has further applications in Radio Astronomy and could help train more researchers in INAOE in heterodyne instrumentation technology.

**Competing interest declaration**

The authors affirm that they have no conflicts of interest to disclose.

**Acknowledgments**

We are grateful to Eimear Gallagher, Brian Ellison, Edgar Colín, Nart Daghestani and Jacob Baars. Special thanks to Dr. Daghestani for allowing Paulina Carmona to visit the Rutherford Appleton Laboratory facilities on February 15, 2024.

[13] Antony Hewish, 'Radioastronomy in Cambridge', in Richard Mason (ed.), Cambridge Minds, Cambridge University Press, 1994, pp. 48-57.

[14] A Brief History of Radio Astronomy in Cambridge. https://www.astro.phy.cam.ac.uk/about/history, accessed [31 May 2024].

[15] Rutherford Appleton Laboratory, https://www.ukri.org/who-we-are/stfc/facilities/rutherford-appleton-laboratory/, accessed [5 June 2024].

[16] Newton Fund, https://www.britishcouncil.org.mx/newton-fund, accessed [7 June 2024].

[17] Paulina Carmona Rodríguez, 'Cooperación internacional entre México y Reino Unido para la creación e implementación del experimento CHARM en el Gran Telescopio Milimétrico', Thesis in Physics (M. P. Ramos-Lara Advisor), Faculty of Science (UNAM) 2024.

[18] They are Dr. Eimear Gallagher, a former doctoral candidate at the International Max Planck Research School on Astrophysics (IMPRS); Dr. Brian Ellison, former Director of the Millimeter Wavelength Group of the Rutherford Appleton Laboratory; Dr. Edgar Colín from Dukosi (Edinburgh); Dr. Nart Daghestani, Research Collaborator of the Millimeter Wavelength Group of the Rutherford Appleton Laboratory; and Dr. Jacob Baars, a former researcher at IMPRS.

[19] International Symposium on Space Terahertz Technology, https://almascience.eso.org/naasc-news/international-symposium-on-space-terahertz-technology-on-april-7-11-2024, accessed [14 June 2024].

[20] Terahertz and Millimetre-Wave Receivers and Radiometers, Future satellite missions, Breadboard demonstrator of the RAL Subharmonic Image Reject Mixer (SHIRM),